\documentclass[twocolumn,showpacs,aps,pra]{revtex4-2}

\usepackage{amsmath,amsfonts,amssymb}
\usepackage{wrapfig}
\usepackage{graphicx}
\usepackage{bbm}
\usepackage{color}
\usepackage{tensor}
\newcommand\bra[1]{\left\langle{#1}\right|}
\newcommand\ket[1]{\left|{#1}\right\rangle}
\usepackage{url}
\usepackage{enumitem}

\usepackage{cancel}
\usepackage{ulem}

\usepackage[colorlinks]{hyperref}

\usepackage{amsthm}
\theoremstyle{definition}

\begin{document}
\title{Quantifying entanglement preservability of experimental processes}
\author{Shih-Hsuan Chen$^{1,2}$}
\author{Meng-Lok Ng$^{1}$}
\author{Che-Ming Li$^{1,2,3,}$}
\email{cmli@mail.ncku.edu.tw}

\affiliation{$^{1}$Department of Engineering Science, National Cheng Kung University, Tainan 70101, Taiwan}
\affiliation{$^{2}$Center for Quantum Frontiers of Research and Technology, National Cheng Kung University, Tainan 70101, Taiwan}
\affiliation{$^{3}$Center for Quantum Technology, Hsinchu, 30013, Taiwan}

\begin{abstract}
Preserving entanglement is a crucial dynamical process for entanglement-based quantum computation and quantum-information processes, such as one-way quantum computing and quantum key distribution. However, the problem of quantifying the ability of an experimental process to preserve two-qubit entanglement in experimentally feasible ways is not well understood. Accordingly, herein, we consider the use of two measures, namely composition and robustness, for quantitatively characterizing the ability of a process to preserve entanglement, referred to henceforth as \textit{entanglement preservability}. A fidelity benchmark is additionally derived to identify the ability of a process to preserve entanglement. We show that the measures and introduced benchmark are experimentally feasible and require only local measurements on single qubits and preparations of separable states. Moreover, they are applicable to all physical processes that can be described using the general theory of quantum operations, e.g., qubit dynamics in photonic and superconducting systems. Our method extends the existing tools for analyzing channels, e.g., channel resource theory, to quantify entanglement preservability for non-trace-preserving quantum processes. The results are of significant interest for applications in quantum-information processing in which entanglement preservation is required.
\end{abstract}

\maketitle

\section{INTRODUCTION}
Quantum entanglement \cite{Horodecki09} is a characteristic of quantum states that describes a collective feature of composite systems and cannot be revealed by the partial trace of the constitutive subsystems. Quantum entanglement is an important physical resource for a variety of quantum-information processing tasks, from remote state preparation \cite{Bennett01,Peters05,Barreiro10} to quantum teleportation \cite{Bennett93,Barrett04,Sherson06,Xia18}, and quantum key distribution \cite{Ekert91,Acin07,Scarani09} to one-way quantum computing \cite{Raussendorf01,Raussendorf03,Walther05}. The performance of such tasks is critically dependent on the quality of entanglement. Thus, identifying entanglement is an essential component in examining the faithfulness of quantum-information processes \cite{Cavalcanti17}. Entanglement identification hinges on whether an entangled state can be produced and preserved well in an experimental process. For example, the fusion of photon pairs \cite{Pan98,Pan12} is an important method for creating entangled photon pairs and genuinely multipartite entangled photons for quantum-information tasks \cite{Zhao04,Chen07}. Ultimately, the faithfulness of these tasks relies on the photon fusion process used to produce entanglement.

Existing methods for identifying whether a given quantum process or channel can create entangled states, i.e., possesses an entangling capacity \cite{Campbell10}, quantitatively characterize the manner in which maximum entanglement is produced from separable states. However, while the ability to create entanglement is clearly important, whether entangled states can be preserved during an experimental process for quantum tasks is also significant since in general quantum-information protocols \cite{Bennett01,Peters05,Barreiro10,Bennett93,Barrett04,Sherson06,Xia18,Ekert91,Acin07,Scarani09,Raussendorf01,Raussendorf03,Walther05,Flamini19}, entanglement must be coherently manipulated, e.g., entanglement swapping between quantum nodes \cite{Goebel09}, or the storage of entanglement in quantum memory \cite{Brennen15}.

Entanglement preservation is a matter of great importance for entanglement physics and its practical applications. Consequently, various methods for investigating such a process characteristic have been proposed. Witnessing non-entanglement-breaking quantum channels \cite{Mao20,Zhen20,Moravcikova10}, for example, provides one approach for confirming entanglement preservation. However, these witnesses do not allow the ability of the process to preserve entangled states to be quantitatively described.

Existing methods for quantifying the characteristics of quantum channels or operations can be broadly classified into two types, namely methods based on deduction \cite{Hsieh17,Kuo19}, and methods based on channel resource theory \cite{Gour19,Liu20,Hsieh19,Saxena20,Rosset18,Theurer19,Liu19,Takagi20,Yuan19,Takagi19,Uola20}, where the latter methods are extended from resource theories for quantum states \cite{Chitambar19}. Such theories consist of three components, namely resource (e.g., entanglement), free states (e.g., separable states), and free operations, i.e., quantum operations that cannot generate resource states from free states and make free states remain free states. Drawing on these resource theories, channel resource theories classify processes into free processes and resource processes, respectively, and define free super-operations that map free processes to free processes. Channel resource theories quantify the channel resources in a completely positive (CP) and trace-preserving (TP) channel \cite{Gour19,Liu20,Hsieh19,Saxena20} and characterize whether the resources are generated or preserved in the process of interest.

Given a specific process characteristic, or a prescribed process ability (referred to as the quantum process capability (QPC) \cite{Kuo19}, e.g., entanglement creation), deduction methods consider the extent to which the results of a given experiment correspond to the predictions of quantum theory (or classical theory).
QPC theory classifies processes into two groups, namely capable processes and incapable processes. Processes of the former type are capable of showing the quantum-mechanical effect on a system prescribed by the specification. By contrast, incapable processes are unable to meet the specification at all.

When considering how to experimentally implement the methods described above in practical cases, such as entanglement in photonic \cite{Pan98,Pan12} or superconducting systems \cite{You05}, there exist several important distinctions between channel resource theories and the QPC method. For example, QPC theory can be used to analyze not only CPTP processes but also non-TP CP processes \cite{Kuo19,Nielsen00}. By contrast, channel resource theories have not analyzed non-TP processes. Furthermore, QPC theory describes and defines the capability of the whole process to cause quantum-mechanical effects on physical systems as a process characteristic. For example, such a capability of the whole process can be a genuinely quantum characteristic that cannot be described by any classical theories. As introduced in Ref.~\cite{Hsieh17}, one of the classical pictures is defined for the whole dynamical process of classical system, where the system follows the classical realism and its evolution is described by classical transition probabilities. Whether or not a process can be simulated by this classical theory \cite{Hsieh17} is not related to the state resources, and has not been analyzed using channel resource theories. However, such a feature is considered as a process characteristic in QPC theory.

Finally, for channel resources which consider the free operations in state resource theories as free channels, the free channels must be completely free \cite{Gour19}. In other words, the tensor product of a free operation and an identity process must remain a free operation in composite systems since, according to resource theory, an identity process is a free process, and thus cannot generate resources from free states. By contrast, incapable processes and capable processes in QPC theory are not required to satisfy this constraint. That is, an identity process can be defined as being either incapable or capable. For example, an identity process can be defined as a capable process with the ability to preserve coherence \cite{Kuo19}.

Drawing on the principles of channel resource theory, the concept of resource preservability \cite{Takagi19,Hsieh19,Saxena20,Takagi20,Yuan19,Uola20} has been proposed to investigate the ability of an experimental process that cannot generate resources to preserve the resources of quantum states. In resource preservability theory, the identity channel is resourceful, and the free channels of resource preservability are not completely free. The theory of resource preservability requires the free super-channels acting on the main system and ancillary systems to be defined and stipulates that the resource preservability of the experimental process on the main system cannot be increased by the free super-channels. Moreover, the ancillary systems cannot provide additional resource preservability for the experimental process, and the state resources in the ancillary systems cannot be generated by the free super-channels.

\begin{figure}[t]
\includegraphics[width=9cm]{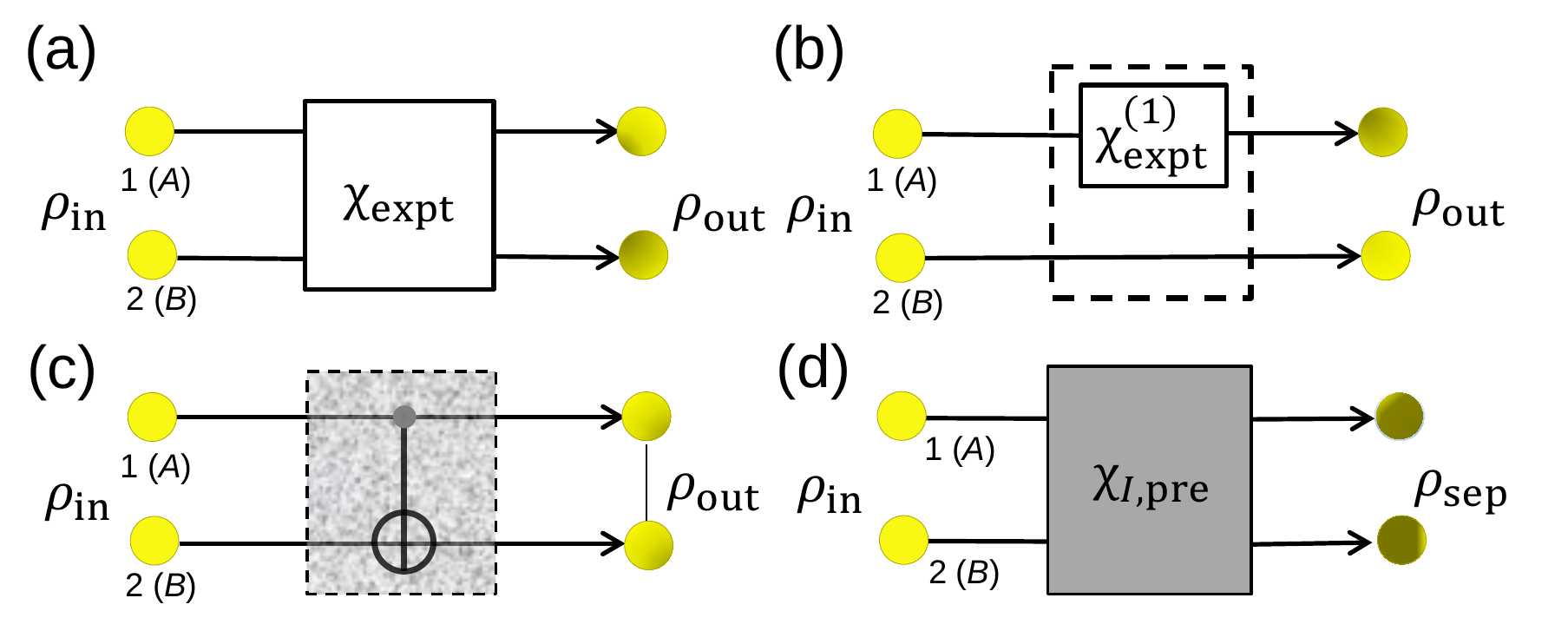}
\caption{Quantifying entanglement preservability of experimental processes. (a) The present study aims to quantify the entanglement preservability of a two-qubit experimental process $\chi_\text{expt}$ on qubits 1 and 2 in subsystems $A$ and $B$, respectively. The process matrix of $\chi_\text{expt}$ can be characterized through analyzing certain output states $\rho_\text{out}$ of specific input sparable states $\rho_\text{in}$ using the process tomography algorithm. See Appendix \ref{process_tomography} for more detailed of process tomography algorithm. (b) For single-qubit processes, denoted as $\chi^{(1)}_\text{expt}$, the \textit{E}-preservability of $\chi^{(1)}_\text{expt}\otimes\chi_I$ represents the ability of a single-qubit process $\chi^{(1)}_\text{expt}$ which acting on subsystem $A$ to preserve entanglement between subsystems $A$ and $B$, where $\chi_I$ represents an identity unitary transformation. (c) \textit{E}-preservability measures can be used to quantify entanglement preservability of the interaction process between two qubits in an experiment such as a controlled-\textsc{not} (CNOT) gate or a controlled-$Z$ gate experiment. (d) Incapable processes of entanglement preservability, $\chi_{\mathcal{I},\text{pre}}$, cannot generate entanglement and makes all input states become separable states, denoted as $\rho_\text{sep}$. With $\chi_{\mathcal{I},\text{pre}}$, we can quantify \textit{E}-preservability of an experimental process $\chi_\text{expt}$ through the method shown in Sec.~\ref{measures}.}
\label{main_concept}
\end{figure}

To quantify the resource preservability of an experimental process in the main system, both the super-channel and the ancillary system must be optimized such that the output state shows the clearest difference between the experimental process and the free super-channels.
Compared to traditional resource preservability theory as will be shown below, the QPC method does not require the use of ancillary systems and is not limited to the quantification of processes having no ability to generate resource. Rather, it follows the quantum operations formalism and characterizes the experimental process by quantum process tomography (QPT) \cite{Nielsen00,Chuang97}, which is experimentally feasible and requires the input of just certain separable states in order to acquire full knowledge of the experimental process, as shown in Fig.~\ref{main_concept}(a).

\begin{table*}[t]
\caption{Comparison between channel resource theory and QPC theory. The differences between channel resource theories and QPC theory are summarized in the three main points. First, they have different purposes. Channel resource theory aims to be constructed as a mathematical structure for channel resource. By contrast, QPC theory aims to quantify the QPC of an experimental process based on the primitive measurement results of the output states. Second, channel resource theories consider free super-channels to describe that resource is non-increasing through free operation. Finally, QPC theory can be used to analyze non-TP processes \cite{Kuo19} that are not considered in channel resource theory, such as the process of generating entangled photon pairs via photon fusion~\cite{Pan98,Pan12}.}
\begin{ruledtabular}
\begin{tabular}{ l c c c }
 & Channel resource theories \cite{Takagi19,Hsieh19,Saxena20,Takagi20,Yuan19,Uola20} & QPC theory \cite{Kuo19} \\ \hline
 1. Purpose &  a formalism constructed for channel resource &  a method for quantifying process capability \\
 2. Definition & resource channel, free channel, and free super-channels & capable process and incapable process\\
 3. Application & CPTP processes & TP and non-TP CP processes\\
\end{tabular}
\end{ruledtabular}
\label{tabletwo}
\end{table*}

The differences between channel resource theories and QPC theory which stem from the different purposes, definitions, and applications of the two theories are summarized in Table~\ref{tabletwo}. In a broad sense, it has been shown that resource theory can be simply a way to divide objects into free and non-free sets~\cite{Gonda19}. There are no assumptions about the type of objects whether the objects are channels or processes. Therefore, QPC theory can be considered as a type of resource theory. As discussed above, while there exists differences between QPC theory and channel resource theory, it is objective to see that they belong to resource theory. Comparing QPC theory with channel resource theories can be helpful to understand the relationship between them.

Driven by the desire to quantify the preservation of entanglement \cite{Kuo19,Hsieh19,Saxena20} during manipulation in practical experiments, we present herein a method for quantifying the composition and robustness of a process as a means of characterizing the capability of the process to preserve entanglement, referred to as the \textit{entanglement preservability capability}, or \textit{E}-preservability for short. The method is based on QPC theory and considers entanglement preservability as a QPC, which completes QPC theory for quantifying both entanglement creation {\cite{Kuo19}} and entanglement preservation.

The considered quantifiers of \textit{E}-preservability satisfy the axioms required for a proper measure \cite{Gour19,Liu19,Liu20} to faithfully quantify the \textit{E}-preservability of a process. Moreover, compared to resource preservability \cite{Hsieh19}, which focuses on processes which cannot generate entanglement, the measures proposed in this work can be used for all two-qubit processes including local single-qubit processes [Fig.~\ref{main_concept}(b)] and two-qubit interaction processes [Fig.~\ref{main_concept}(c)]. (Note that a detailed comparison of the proposed \textit{E}-preservability and the quantifiers derived from channel resource theory \cite{Yuan19,Hsieh19} is presented in Appendix~\ref{ComparisonRT} of this paper.) Two concrete methods for measuring the entanglement preservability are presented. Moreover, a process-fidelity criterion for determining whether a process can preserve entanglement is also provided. We demonstrate the application of the introduced measures and entanglement preservation criterion to two practical systems widely used in entanglement experiments. While creating and preserving entanglement are essential processes for quantum information, the relationship between them is unclear. Thus, we further reveal their relationship and present three concrete examples to compare the quantification of entanglement preservation with the capability of entanglement creation \cite{Kuo19}.

\section{Entanglement preservability}
\subsection{Capable processes, incapable processes, and \textit{E}-preservability}

To quantify the entanglement preservability of an experimental process, we treat \textit{E}-preservability as a QPC. Moreover, we use process matrix $\chi$, a positive Hermitian matrix from the QPT algorithm, to describe a process of the mapping from the input state $\rho_{\text{in}}$ to the output state $\rho_{\text{out}}=\chi(\rho_{\text{in}})$. (See Appendix \ref{process_tomography} for a detailed review of process tomography and the process matrix, including the derivation of $\chi$ from experimentally measurable quantities.) We further use QPC theory to classify two-qubit processes into capable and incapable processes.\\

\begin{itemize}[labelsep = 2 pt,leftmargin = 0 pt,topsep = 2 pt,parsep = 0 pt]

\item[]\textbf{Definition. Incapable process and capable process for \textit{E}-preservability.} A process is said to be an \textit{incapable} process which does not possess \textit{E}-preservability, denoted as $\chi_{\mathcal{I},\text{pre}}$, if it makes all input states become separable states. A process is then said to be a \textit{capable} process which has \textit{E}-preservability if the process cannot be described by $\chi_{\mathcal{I},\text{pre}}$ at all.\\

\end{itemize}

\subsection{Properties of incapable processes}
According to the definition, an incapable process, $\chi_{\mathcal{I},\text{pre}}$, makes all states separable, including entangled states. Since an incapable process destroys entanglement between two subsystems in the main system interacting with environment, an incapable process can be considered as an entanglement-annihilating channel \cite{Moravcikova10} which is different from entanglement-breaking channel that destroy entanglement between the main system and environment. (See Appendix~\ref{ComparisonRT} for a concrete example of entanglement-breaking channel.) Moreover, an incapable process should remain incapable following manipulation by another $\chi_{\mathcal{I},\text{pre}}$. In other words, an incapable process possesses the following properties:\\

\begin{itemize}[labelsep = 2 pt,leftmargin = 21 pt,topsep = 2 pt,parsep = 0 pt]
\item[(P1)] If a process is composed of two cascaded incapable processes, i.e., $\chi=\chi_{\mathcal{I},\text{pre}1}\circ\chi_{\mathcal{I},\text{pre}2}$, then the resulting process, $\chi$, is also incapable, where $\circ$ denotes the concatenation operator. The process $\chi$ is incapable since after $\chi_{\mathcal{I},\text{pre}1}$ makes the input states separable, the output states of these states after $\chi_{\mathcal{I},\text{pre}2}$ are also separable.
\item[(P2)] If a process is a linear combination of incapable processes, i.e., $\chi$=$\sum_{n} p_{n}\chi_{\mathcal{I},\text{pre}n}$, where $\sum_{n} p_{n}=1$, then $\chi$ is also an incapable process since the mixtures of separable states are also separable states.\\
\end{itemize}
As will be seen in Sec.~\ref{MP}, these two properties are used to show that the introduced measures can faithfully quantify the \textit{E}-preservability of a process.

\subsection{Constructing incapable processes $\chi_{\mathcal{I},\text{pre}}$}\label{properties}
In practical experiments, if we want to examine an experimental process through comparing which with incapable processes $\chi_{\mathcal{I},\text{pre}}$, we first need to describe incapable processes $\chi_{\mathcal{I},\text{pre}}$ in terms of process matrices. In this subsection, we will show how to describe and construct the process matrices of incapable processes through the definition and the properties of incapable processes. We first start with the property of output states from incapable processes $\chi_{\mathcal{I},\text{pre}}$. Since the output states of $\chi_{\mathcal{I},\text{pre}}$ must be separable, denoted by $\rho_\text{sep}$ [Fig.~\ref{main_concept}(d)], the partial transposition (PT) of $\rho_\text{sep}$, denoted as $\rho_\text{sep}^{\text{PT}}$, must be positive semi-definite, i.e., $\rho_\text{sep}^{\text{PT}}\geq0$, which is known as the positive partial transpose (PPT) criterion \cite{Peres96,Horodecki96}.

With the property of output state from $\chi_{\mathcal{I},\text{pre}}$, we then show how to construct the process matrix $\chi_{\mathcal{I},\text{pre}}$ through the QPT algorithm. In QPT algorithm, a process matrix of a two-qubit system can be obtained by inputing 16 linear independent and experimentally preparable input states, $\rho'_{\text{in},m}$, and analyzing the corresponding density matrices of the output states, $\rho'_{\text{out},m}$.
See Appendix \ref{process_tomography} for the details. Through the 16 output states $\rho'_{\text{out},m}$, one is able to construct a process matrix $\chi_{\mathcal{I},\text{pre}}$ that satisfies the definition of incapable processes. In the following, we will show that as the 16 output states for constructing $\chi_{\mathcal{I},\text{pre}}$ are separable, the output of an arbitrary input state through the process is a separable state.
To show the proof of the above statement, we first consider an arbitrary input two-qubit state of the form:
\begin{eqnarray}
\rho_{\text{in}}
&=&\left[\begin{array}{cccc}a_1 & a_2 & a_3  & a_4  \\a_5 & a_6 & a_7  & a_8  \\a_9 & a_{10} & a_{11}  & a_{12}  \\a_{13} & a_{14} & a_{15}  & a_{16} \end{array}\right],\label{inputstate}
\end{eqnarray}
where $a_i$, for $i=1,2,...,16$ are the matrix elements of the density matrix $\rho_{\text{in}}$. Since $\rho_{\text{in}}$ (\ref{inputstate}) can be represented as a linear combination of the 16 independent density matrices $\rho'_{\text{in},m}$, the corresponding output states, $\rho_{\text{out}}=\chi_{\mathcal{I},\text{pre}}(\rho_{\text{in}})$, can be represented by the outputs of the 16 input states, $\rho'_{\text{out},m}$, as well. That is,
\begin{eqnarray}
\rho_{\text{out}}&=&(a_1-2a_2-2a_3-2a_4)\rho'_{\text{out},00}\nonumber\\
&&+(a_6-2a_5-2a_7-2a_8)\rho'_{\text{out},01}\nonumber\\
&&+(a_{11}-2a_9-2a_{10}-2a_{12})\rho'_{\text{out},10}\nonumber\\
&&+(a_{16}-2a_{13}-2a_{14}-2a_{15})\rho'_{\text{out},11}\nonumber\\ &&+(a_2+a_5)\rho'_{\text{out},0+}+(a_2-a_5)\rho'_{\text{out},0R}\nonumber\\
&&+(a_3+a_9)\rho'_{\text{out},+0}+(a_3-a_9)\rho'_{\text{out},R0}\nonumber\\
&&+(a_4+a_{13})\rho'_{\text{out},\phi^+}+(a_4-a_{13})\rho'_{\text{out},\phi^{+i}}\nonumber\\
&&+(a_7+a_{10})\rho'_{\text{out},\psi^+}+a_7-a_{10})\rho'_{\text{out},\psi^{+i}}\nonumber\\
&&+(a_{8}+a_{14})\rho'_{\text{out},1+}+(a_{8}-a_{14})\rho'_{\text{out},R1}\nonumber\\
&&+(a_{12}+a_{15})\rho'_{\text{out},1+}+(a_{12}-a_{15})\rho'_{\text{out},1R},\label{outputstates}
\end{eqnarray}
where the details of $\rho'_{\text{out},m}$ are shown in Appendix \ref{process_tomography}. Then, we apply the partial transposition to $\rho_{\text{out}}$ (\ref{outputstates}). The resulting $\rho_{\text{out}}^{\text{PT}}$ can be represented as a linear combination of the 16 output states after the partial transposition, $\rho'^{\text{PT}}_{\text{out},m}$, i.e.,
\begin{eqnarray}
\rho_{\text{out}}^{\text{PT}}&=&(a_1-2a_2-2a_3-2a_4)\rho'^{\text{PT}}_{\text{out},00}\nonumber\\
&&+(a_6-2a_5-2a_7-2a_8)\rho'^{\text{PT}}_{\text{out},01}\nonumber\\
&&+(a_{11}-2a_9-2a_{10}-2a_{12})\rho'^{\text{PT}}_{\text{out},10}\nonumber\\
&&+(a_{16}-2a_{13}-2a_{14}-2a_{15})\rho'^{\text{PT}}_{\text{out},11}\nonumber\\ &&+(a_2+a_5)\rho'^{\text{PT}}_{\text{out},0+}+(a_2-a_5)\rho'^{\text{PT}}_{\text{out},0R}\nonumber\\
&&+(a_3+a_9)\rho'^{\text{PT}}_{\text{out},+0}+(a_3-a_9)\rho'^{\text{PT}}_{\text{out},R0}\nonumber\\
&&+(a_4+a_{13})\rho'^{\text{PT}}_{\text{out},\phi^+}+(a_4-a_{13})\rho'^{\text{PT}}_{\text{out},\phi^{+i}}\nonumber\\
&&+(a_7+a_{10})\rho'^{\text{PT}}_{\text{out},\psi^+}+(a_7-a_{10})\rho'^{\text{PT}}_{\text{out},\psi^{+i}}\nonumber\\
&&+(a_{8}+a_{14})\rho'^{\text{PT}}_{\text{out},1+}+(a_{8}-a_{14})\rho'^{\text{PT}}_{\text{out},R1}\nonumber\\
&&+(a_{12}+a_{15})\rho'^{\text{PT}}_{\text{out},1+}+(a_{12}-a_{15})\rho'^{\text{PT}}_{\text{out},1R}.\label{PTstate}
\end{eqnarray}

Since $\rho'_{\text{out},m}$ (\ref{outputstates}) are separable states, $\rho'^{\text{PT}}_{\text{out},m}$ (\ref{PTstate}) are positive operators, i.e., $\rho'^{\text{PT}}_{\text{out},m}\geq 0$ \cite{Peres96,Horodecki96}. Thus, the 16 $\rho'^{\text{PT}}_{\text{out},m}$ in Eq.~(\ref{PTstate}) can be considered as the outputs of a CP process, denoted by $\chi_{\text{sep}}$, and its process matrix can be obtained by using the QPT algorithm \cite{Nielsen00} as shown in Appendix \ref{process_tomography}. As the input state is in the form of Eq.~(\ref{inputstate}), the corresponding output would be $\rho_{\text{out}}^{\text{PT}}$ (\ref{PTstate}), i.e., $\chi_{\text{sep}}(\rho_{\text{in}})=\rho_{\text{out}}^{\text{PT}}$. Since $\chi_{\text{sep}}$ is a CP map for an arbitrary two-qubit input state $\rho_{\text{in}}$, it is certainly the case that $\rho_{\text{out}}^{\text{PT}}$ (\ref{PTstate}) are positive semi-definite, i.e., $\rho_{\text{out}}^{\text{PT}}\geq 0$, which implies that $\rho_{\text{out}}$ (\ref{outputstates}) is separable for an arbitrary input state $\rho_{\text{in}}$ (\ref{inputstate}). This proves that given the 16 separable states $\rho'_{\text{out},m}$ for constructing the process matrix of $\chi_{\mathcal{I},\text{pre}}$, all the output states of $\chi_{\mathcal{I},\text{pre}}$ for arbitrary input states are separable.

\subsubsection{Describing incapable processes $\chi_{\mathcal{I},\text{pre}}$ in semi-definite programming.}\label{SecSDP}
To quantitatively compare an experimental process with incapable processes, we analyze the experimental process via semi-definite programming (SDP) \cite{Lofberg,sdpsolver}. First of all, we need to describe $\chi_{\mathcal{I},\text{pre}}$ in SDP according to the method introduced in Sec.~\ref{properties}. In SDP, $\chi_{\mathcal{I},\text{pre}}$ is determined by a concrete set of constraints on $\chi_{\mathcal{I},\text{pre}}$, denoted by $D(\tilde{\chi}_{\mathcal{I},\text{pre}})$, which is formulated as follows:
\begin{eqnarray}\label{Iconstraints}
&&\chi_{\mathcal{I},\text{pre}}\geq0,\nonumber\\
&&\rho'_{\text{out},m}=\chi_{\mathcal{I},\text{pre}}(\rho'_{\text{in},m})\geq0,\\ &&\rho'^{\text{PT}}_{\text{out},m}=\chi_{\mathcal{I},\text{pre}}(\rho'_{\text{in},m})^{\text{PT}}\geq0 \ \   \forall  \rho'_{\text{in},m}.\nonumber
\end{eqnarray}
The first constraint ensures that $\chi_{\mathcal{I},\text{pre}}$ is a process matrix of a CP map. The next constraint states that all the 16 output states $\rho'_{\text{out},m}$ for QPT are positive semi-definite. The last constraint states that all these output states are separable examined by using the PPT criterion. As will be shown in Sec. \ref{measures}, the incapable processes described in the SDP programing are helpful to quantify \textit{E}-preservability for experimental processes.


\subsection{Examples of capable and incapable processes}\label{example}
\subsubsection{Incapable process}
Here, we provide two concrete examples to demonstrate incapable processes $\chi_{\mathcal{I},\text{pre}}$ and capable processes introduced above.

Since quantum operations consisting of local operations and shared randomness (LOSR) \cite{Rosset20,Miguel20} preserve the separability of separable states, we first consider a mixture of quantum operations on individual subsystems denoted as $\chi_{\text{LOSR}}$, which is of the form:
\begin{equation}\label{ChiPreI}
\chi_{\text{LOSR}}=\sum_{i}p_i\chi_{i}^{A}\otimes\chi_{i}^{B},
\end{equation}
where $\chi_{i}^{A}$ and $\chi_{i}^{B}$ are the quantum operations on each qubit, and $p_i$ is the probability distribution of $\chi_{i}^{A}$ and $\chi_{i}^{B}$. (See Ref.~\cite{example_LOCC} as a concrete example realizing Eq.~(\ref{ChiPreI}) in a scenario consisting of local operations and classical communications (LOCC).) Since incapable processes make all input states, including not only separable states but also entangled states, become separable, $\chi_{\text{LOSR}}$ is incapable only when which can make all of the entangled inputs become separable states. Through the methods to construct $\chi_{\mathcal{I},\text{pre}}$ presented in Sec.~\ref{properties}, if the outputs of the four maximally entangled inputs: $\rho'_{\text{out},\phi^+}$, $\rho'_{\text{out},\phi^{+i}}$, $\rho'_{\text{out},\psi^+}$, and $\rho'_{\text{out},\psi^{+i}}$ are separable states, then $\chi_{\text{LOSR}}$ is an incapable process.

It is worth noting that, through the Choi-Jamio\l kowski isomorphism \cite{Jamiolkowski72,Choi75,Jiang13}, the output states of the maximally entangled input states, denoted as $\rho_\text{ent}$, after manipulation by $\chi_{\mathcal{I},\text{pre}}$ with the above representation in Eq.~(\ref{ChiPreI}) can be written as
\begin{equation}
\sum_{i}p_i\chi_{i}^{A}\otimes\chi_{i}^{B} (\rho_\text{ent})=\chi'_{\mathcal{I}}\otimes\chi_I (\rho_\text{ent})\nonumber,
\end{equation}
where $\chi'_{\mathcal{I}}=\sum_{i}p_i\chi_{i}^{A}\circ(\chi_{i}^{B})^T$ and $\chi_{I}$ is an identity unitary transformation. Thus, the output states of incapable processes $\chi_{\mathcal{I},\text{pre}}$ can be represented as
\begin{equation}\label{PreI}
\chi_{\text{LOSR}}(\rho_\text{ent})=\chi'_{\mathcal{I}}\otimes\chi_I (\rho_\text{ent}).
\end{equation}
This helps identify capable and incapable processes for single-qubit processes of preserving entanglement. A single-qubit process is capable (incapable) to preserve entanglement if the tensor product of the single-qubit process and an identity unitary transformation is a two-qubit capable (incapable) process as shown in Sec.~\ref{SecIBMQ}. This example of illustrating how to identify incapable processes consisting of local operations is also considered in Ref.~\cite{Yuan19}. See Appendix~\ref{ComparisonRT} for a more detailed discussion.

\subsubsection{Capable process}
A capable processes with \textit{E}-preservability is defined to be the processes that cannot be described by $\chi_{\mathcal{I},\text{pre}}$. Here, let us consider a process $\chi_{\text{expt}}$ with the following effect on input states:
\begin{eqnarray}\label{CExample}
\chi_{\text{expt}}(\rho)&=& \text{tr}(\ket{00}\!\!\bra{00}\rho)\ket{\phi^+}\!\!\bra{\phi^+}  \nonumber\\
&&+\text{tr}\left[(I-\ket{00}\!\!\bra{00})\rho\right]\frac{I-\ket{\phi^+}\!\!\bra{\phi^+}}{3},
\end{eqnarray}
where $I$ is identity matrix and $\ket{\phi^+}=(\ket{00}+\ket{11})/\sqrt{2}$. The process $\chi_{\text{expt}}$ shows the ability to destroy entanglement, e.g.,
\begin{equation}
\chi_{\text{expt}}(\ket{\phi^+}\!\!\bra{\phi^+})=\frac{1}{6}\ket{\phi^+}\!\!\bra{\phi^+}+\frac{5I}{24}, \nonumber
\end{equation}
and $\chi_{\text{expt}}(\ket{\phi^+}\!\!\bra{\phi^+})^{\text{PT}}\geq 0$.
However, it is also possible to see entangled output states given specific inputs, e.g.,
\begin{equation}
\chi_{\text{expt}}(\ket{00}\!\!\bra{00})=\ket{\phi^+}\!\!\bra{\phi^+}.\nonumber
\end{equation}
Therefore, the $\chi_{\text{expt}}$ cannot be described by any incapable processes and thus shows the capability to preserve entanglement. After identifying \textit{E}-preservability, we will further illustrate how to quantify \textit{E}-preservability of $\chi_{\text{expt}}$ in the next section.

\section{E-preservability measures}\label{measures}
Having introduced the definitions and properties of capable processes and incapable processes $\chi_{\mathcal{I},\text{pre}}$ above, we now investigate how to construct a faithful measure, denoted as $C(\chi_{\text{expt}})$, for a given experimental process matrix $\chi_{\text{expt}}$ which can be obtained by the preparation of specific separable states and local measurements in practical experiments. We illustrate two measures, namely the \textit{E}-preservability composition, $\alpha_\text{pre}$, and the \textit{E}-preservability robustness, $\beta_\text{pre}$, for quantifying the \textit{E}-preservability of $\chi_{\text{expt}}$.

\subsection{\textit{E}-preservability composition $\alpha_\text{pre}$}\label{composition}
Any experimental process, $\chi_{\text{expt}}$, can be represented as a linear combination of capable and incapable processes \cite{Hsieh17,Kuo19}, i.e.,
\begin{equation}
\chi_{\text{expt}}=a\chi_{\mathcal{C},\text{pre}}+(1-a)\chi_{\mathcal{I},\text{pre}},\label{com}
\end{equation}
where $a\geq0$ and $\chi_{\mathcal{C},\text{pre}}$ are capable processes with $a=1$. The \textit{E}-preservability composition of $\chi_{\text{expt}}$ can then be defined as
\begin{equation}
\alpha_\text{pre}\equiv\min_{\chi_{\mathcal{I},\text{pre}}}a,\label{COM}
\end{equation}
where $\alpha_\text{pre}$ specifies the minimum amount of capable process $\chi_{\mathcal{C},\text{pre}}$ that can be found in the experimental process.

In practical experiments on entanglement, after obtaining the process matrix $\chi_{\text{expt}}$ of the experimental process via the QPT algorithm, the \textit{E}-preservability composition, $\alpha_\text{pre}$, of $\chi_{\text{expt}}$ can be obtained by minimizing the following quantity via SDP \cite{Lofberg,sdpsolver} with MATLAB:
\begin{equation}
\alpha_\text{pre}=\min_{\tilde{\chi}_{\mathcal{I},\text{pre}}}[1-\text{tr}(\tilde{\chi}_{\mathcal{I},\text{pre}})].
\end{equation}
Note that the solution is obtained under a set of specified conditions such that $\chi_{\text{expt}}-\tilde{\chi}_{\mathcal{I},\text{pre}}=\tilde{\chi}_{\mathcal{C},\text{pre}}\geq0$ and constraints for the incapable process $D(\tilde{\chi}_{\mathcal{I},\text{pre}})$ in Eq.~(\ref{Iconstraints}). Here, $\tilde{\chi}_{\mathcal{I},\text{pre}}$ and $\tilde{\chi}_{\mathcal{C},\text{pre}}$ are both unnormalized process matrices and possess the properties $\text{tr}(\tilde{\chi}_{\mathcal{I},\text{pre}})=\text{tr}((1-a)\chi_{\mathcal{I},\text{pre}})=1-a$ and $\text{tr}(\tilde{\chi}_{\mathcal{C},\text{pre}})=\text{tr}(a\chi_{\mathcal{C},\text{pre}})=a$, respectively. Furthermore, $D(\tilde{\chi}_{\mathcal{I},\text{pre}})$ places constraints on the process matrix construction of the incapable process in the QPT algorithm as shown in Eq.~(\ref{Iconstraints}), which specify how the input and output states for the QPT algorithm should behave under the incapable process, $\tilde{\chi}_{\mathcal{I},\text{pre}}$. For the example of the capable process in Eq.~(\ref{CExample}), the \textit{E}-preservability composition $\alpha_\text{pre}$ of $\chi_{\text{expt}}$ is $0.1333$.

It is worth noting that although the 16 input states $\rho'_{\text{in},m}$, including 4 entangled states and 12 separable states, are used in the QPT algorithm for Eq.~(\ref{outputstates}) and the related discussions (see Appendix \ref{process_tomography} for details), one can use only separable states to obtain the process matrix $\chi_{\text{expt}}$ in practical experiments. For example, we can use the tensor products of the identity matrix and the three Pauli matrices as the 16 matrix elements for a two-qubit density matrix. Thus, the input states chosen for the QPT algorithm can be represented as a linear combination of the tensor products of the eigenstates of the three Pauli matrices. These 36 input states are all separable.

\subsection{\textit{E}-preservability robustness $\beta_\text{pre}$}
An experimental process, $\chi_{\text{expt}}$, can become incapable by mixing with noise, i.e.,
\begin{equation}
\frac{\chi_{\text{expt}}+b\chi'}{1+b}=\chi_{\mathcal{I},\text{pre}},\label{rob}
\end{equation}
where $b\geq 0$ and $\chi'$ is the noise process. The \textit{E}-preservability robustness of $\chi_{\text{expt}}$ is defined as the minimum amount of noise which must be added to $\chi_{\text{expt}}$ such that $\chi_{\text{expt}}$ becomes $\chi_{\mathcal{I},\text{pre}}$, i.e.,
\begin{equation}
\beta_\text{pre}\equiv\min_{\chi'}b,\label{ROB}
\end{equation}
and can be obtained by using SDP to solve
\begin{equation}
\beta_\text{pre}=\min_{\tilde{\chi}_{\mathcal{I},\text{pre}}}[\text{tr}(\tilde{\chi}_{\mathcal{I},\text{pre}})-1],
\end{equation}
subject to the conditions $\text{tr}(\tilde{\chi}_{\mathcal{I},\text{pre}})\geq1,\ \tilde{\chi}_{\mathcal{I},\text{pre}}-\chi_{\text{expt}}\geq0$, and the constraints for incapable processes $D(\tilde{\chi}_{\mathcal{I,\text{pre}}})$ given in Eq.~(\ref{Iconstraints}), where $\text{tr}(\tilde{\chi}_{\mathcal{I},\text{pre}})=\text{tr}((1+b)\chi_{\mathcal{I},\text{pre}})=1+b$. Note that the constraints $\text{tr}(\tilde{\chi}_{\mathcal{I},\text{pre}})\geq1,\ \tilde{\chi}_{\mathcal{I},\text{pre}}-\chi_{\text{expt}}\geq0$ ensure that $\beta_\text{pre}\geq0$ and $\chi'$ is positive semi-definite, respectively. It is worth noting that the concept
of robustness has been widely used in previous studies to quantify processes with respect to different characteristics \cite{Hsieh17,Rosset18,Hsieh19,Kuo19,Takagi19,Liu19}.

\subsection{$\alpha_\text{pre}$ and $\beta_\text{pre}$ are sensible \textit{E}-preservability measures}\label{MP}
The \textit{E}-preservability measures $\alpha_\text{pre}$ and $\beta_\text{pre}$ satisfy the conditions that a proper measure should be satisfied \cite{Gour19,Liu19,Liu20}. First, for a sensible measure $C(\chi_{\text{expt}})$, the values should be non-negative for all processes and equal to zero for incapable processes. The \textit{E}-preservability measures $\alpha_\text{pre}$ and $\beta_\text{pre}$ of $\chi_{\mathcal{C},\text{pre}}$ confirm that $C(\chi_{\mathcal{C},\text{pre}})>0$. By contrast, the measures of $\chi_{\mathcal{I},\text{pre}}$ are minimum, i.e., $C(\chi_{\mathcal{I},\text{pre}})=0$. Furthermore, the \textit{E}-preservability of a process which incorporates incapable processes $\chi_{\mathcal{I},\text{pre}}$ will not increase since the incapable processes do not have \textit{E}-preservability. Finally, the \textit{E}-preservability of a process incorporating mixed incapable processes $\sum_{n}p_{n}\chi_{\mathcal{I},\text{pre}n}$ will also not increase. Thus, if a measure $C(\chi_{\text{expt}})$ is to faithfully quantify \textit{E}-preservability, it should satisfy the following three conditions:\\
\begin{itemize}[labelsep = 2 pt,leftmargin = 30 pt,topsep = 2 pt,parsep = 0 pt]
\item[(MP1)]Faithfulness: $C(\chi)=0$ if and only if $\chi$ is incapable.
\item[(MP2)]Monotonicity: $C(\chi\circ\chi_{\mathcal{I},\text{pre}})\leq C(\chi)$, the measure of \textit{E}-preservability of a process $\chi$ will not increase following extension with an incapable process.
\item[(MP3)]Convexity: $C(\sum_{n}p_{n}\chi\circ\chi_{\mathcal{I},\text{pre}n})\leq\sum_{n}p_{n}C(\chi\circ\chi_{\mathcal{I},\text{pre}n})$, the mixing of processes will not increase the \textit{E}-preservability.\\
\end{itemize}

\begin{figure*}[t]
\includegraphics[width=18cm]{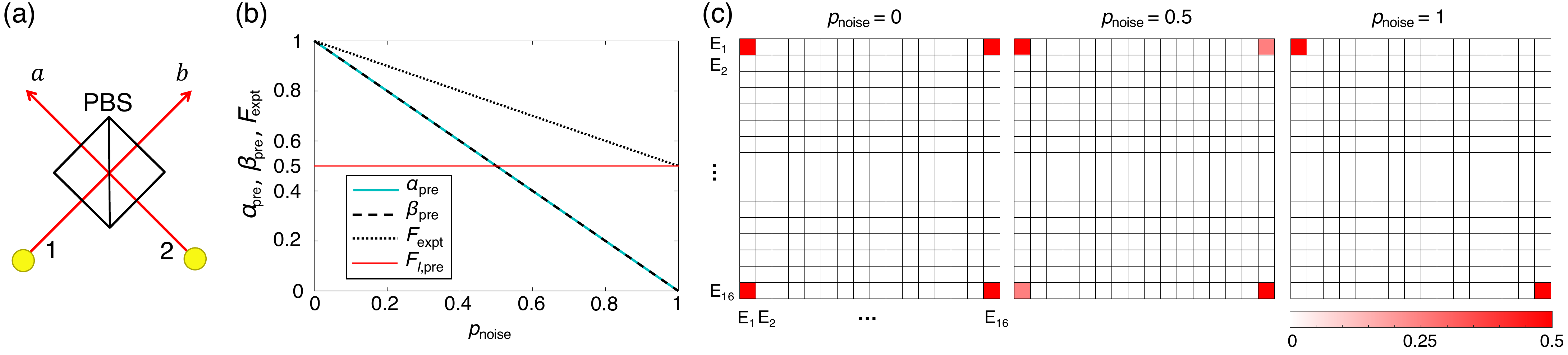}
\caption{Evaluating photon fusion processes using \textit{E}-preservability measures. (a) The fusion of photon pairs can be realized by a PBS through inputting two individual photons in two different spatial modes ($1$ and $2$) simultaneously and then post-selecting the two outputs in different modes ($a$ and $b$). (b) The processes of imperfect fusion with different noise intensity, $p_\text{noise}$, are examined by two \textit{E}-preservability measures, $\alpha_\text{pre}$ and $\beta_\text{pre}$, and the fidelity criterion. The \textit{E}-preservability criterion for photon fusion is $F_{\mathcal{I},\text{pre}}=0.5$. $\chi_\text{expt}$ becomes incapable at $p_\text{noise}=1$. (c) Illustrative process matrices for photon fusion under noise intensities of $p_\text{noise}=0$, $0.5$ and $1$. (See Appendix \ref{process_tomography} for the details of operators $E_k$, $k=1,2,...,16$, and the elements of process matrices.)}
\label{fusion}
\end{figure*}

The \textit{E}-preservability measures $\alpha_\text{pre}$ and $\beta_\text{pre}$ satisfy (MP1) directly according to the definitions of $\chi_{\mathcal{I},\text{pre}}$, $\alpha_\text{pre}$ and $\beta_\text{pre}$, respectively. To prove that $\alpha_\text{pre}$ and $\beta_\text{pre}$ also satisfy (MP2), the process of $\chi$ incorporated with incapable processes $\chi_{\mathcal{I},\text{pre}}$ can be represented in the form shown in Eqs.~(\ref{com}) and (\ref{COM}) as follows:
\begin{eqnarray}
\chi\circ\chi_{\mathcal{I},\text{pre}} &=&(\alpha_\text{pre}\chi_{\mathcal{C},\text{pre}}+(1-\alpha_\text{pre})\chi_{\mathcal{I},\text{pre}})\circ\chi_{\mathcal{I},\text{pre}} \nonumber \\
&=&\alpha_\text{pre}\chi_{\mathcal{C},\text{pre}}\circ\chi_{\mathcal{I},\text{pre}}+(1-\alpha_\text{pre})\chi_{\mathcal{I},\text{pre}}\circ\chi_{\mathcal{I},\text{pre}} \nonumber \\
&=&\alpha_\text{pre}'\chi_{\mathcal{C},\text{pre}}+(1-\alpha_\text{pre}')\chi_{\mathcal{I},\text{pre}}.
\end{eqnarray}
According to property (P1), $\chi_{\mathcal{I},\text{pre}}\circ\chi_{\mathcal{I},\text{pre}}$ must be incapable. Thus, the amount of $\chi_{\mathcal{I},\text{pre}}$ in $\chi\circ\chi_{\mathcal{I},\text{pre}}$ must be greater than $\chi$, i.e., $1-\alpha_\text{pre}'\geq 1-\alpha_\text{pre}$, which implies $\alpha_\text{pre}'\leq \alpha_\text{pre}$ and shows that the \textit{E}-preservability composition $\alpha_\text{pre}$ satisfies (MP2). Since the amount of $\chi_{\mathcal{C},\text{pre}}$ does not increase, the minimum amount of noise, i.e., $\beta_\text{pre}$ in Eq.~(\ref{ROB}), that makes $\chi\circ\chi_{\mathcal{I},\text{pre}}$ become $\chi_{\mathcal{I},\text{pre}}$ also does not increase. Thus, $\beta_\text{pre}$ also satisfies (MP2).

To prove that $\alpha_\text{pre}$ and $\beta_\text{pre}$ satisfy (MP3), we represent $\sum_{n}p_{n}\chi\circ\chi_{\mathcal{I},\text{pre}n}$ as
\begin{eqnarray}
\sum_{n}p_{n}\chi\circ\chi_{\mathcal{I},\text{pre}n} =a'_\text{pre}\chi_{\mathcal{C'},\text{pre}}+(1-a'_\text{pre})\chi_{\mathcal{I'},\text{pre}},\nonumber
\end{eqnarray}
where
\begin{eqnarray}
a'_\text{pre}&=&\sum_{n}p_{n}\alpha_{\text{pre},n}=\sum_{n}p_{n}C(\chi\circ\chi_{\mathcal{I},\text{pre}n}),\nonumber \\ \chi_{\mathcal{C'},\text{pre}}&=&\frac{\sum_{n}p_{n}\alpha_{\text{pre},n}\chi_{\mathcal{C},\text{pre}n}}{a'_\text{pre}}, \nonumber \\ \chi_{\mathcal{I'},\text{pre}}&=&\frac{\sum_{n}p_{n}(1-\alpha_{\text{pre},n})\chi_{\mathcal{I},\text{pre}n}}{1-a'_\text{pre}}. \nonumber
\end{eqnarray}
It is worth noting that $a'_\text{pre}$ is not necessarily the optimal $\alpha_{\text{pre}}=C(\sum_{n}p_{n}\chi\circ\chi_{\mathcal{I},\text{pre}n})$. Since $\chi_{\mathcal{I'},\text{pre}}$ is incapable according to (P2), $\sum_{n}p_{n}\chi\circ\chi_{\mathcal{I},\text{pre}n}$ must contain an amount $(1-a'_\text{pre})$ of incapable process, where $(1-a'_\text{pre})\leq(1-\alpha_\text{pre})$ and $\alpha_\text{pre}\leq a'_\text{pre}$. In other words, $\alpha_\text{pre}$ satisfies (MP3).

To show that $\beta_\text{pre}$ also satisfies (MP3), $\sum_{n}p_{n}\chi\circ\chi_{\mathcal{I},\text{pre}n}$ can be represented as
\begin{eqnarray}
\sum_{n}p_{n}\chi\circ\chi_{\mathcal{I},\text{pre}n} =(1+b'_\text{pre})\chi_{\mathcal{I'},\text{pre}}-b'_\text{pre}\chi''\nonumber
\end{eqnarray}
according to Eq.~(\ref{rob}), where
\begin{eqnarray}
b'_\text{pre}&=&\sum_{n}p_{n}\beta_{\text{pre},n}=\sum_{n}p_{n}C(\chi\circ\chi_{\mathcal{I},\text{pre}n}), \nonumber \\ \chi_{\mathcal{I'},\text{pre}}&=&\frac{\sum_{n}p_{n}(1+\beta_{\text{pre},n})\chi_{\mathcal{I},\text{pre}n}}{1+b'_\text{pre}}, \nonumber \\
\chi''&=&\frac{\sum_{n}p_{n}\beta_{\text{pre},n}\chi''_{n}}{b'_\text{pre}}, \nonumber
\end{eqnarray}
and $\chi''_{n}$ is the noise process for each $\chi\circ\chi_{\mathcal{I},\text{pre}n}$.
Since $\beta_\text{pre}=C(\sum_{n}p_{n}\chi\circ\chi_{\mathcal{I},\text{pre}n})$ is the optimal $b$ in Eq.~(\ref{rob}), i.e. $\beta_\text{pre}\leq b'_\text{pre}$, then $\beta_\text{pre}$ also satisfies (MP3).

If a measure does not satisfy (MP1)-(MP3), the measured value of the \textit{E}-preservability of a process quantified by this measure may increase following manipulation by an incapable processes $\chi_{\mathcal{I},\text{pre}}$. Hence, this measure cannot reliably show \textit{E}-preservability, since \textit{E}-preservability will not increase following incorporation or mixing with incapable processes.

\section{\textit{E}-preservability criterion}
When $\chi_{\text{expt}}$ is created with respect to a target process, $\chi_{\text{target}}$, the similarity between them can be examined using the process fidelity, $F_{\text{expt}}\equiv \text{tr}(\chi_{\text{expt}}\chi_{\text{target}})$. In particular, $\chi_{\text{expt}}$ is judged to have \textit{E}-preservability and to be close to the target process if it goes beyond the best mimicry achieved by incapable processes to $\chi_{\text{target}}$, i.e.,
\begin{equation}
F_{\text{expt}}>F_{\mathcal{I},\text{pre}}\equiv\max_{\chi_{\mathcal{I},\text{pre}}}[\text{tr}(\chi_{\mathcal{I},\text{pre}}\chi_{\text{target}})],\label{F}
\end{equation}
which means that $\chi_{\text{expt}}$ is a faithful operation which cannot be simulated by any incapable processes. Notably, $F_{\mathcal{I},\text{pre}}$ in Eq.~(\ref{F}) can be evaluated by performing the following maximization task with SDP: $F_{\mathcal{I},\text{pre}}=\max_{\tilde{\chi}_{\mathcal{I},\text{pre}}}[\text{tr}(\tilde{\chi}_{\mathcal{I},\text{pre}}\chi_{\text{target}})]$, under $D(\tilde{\chi}_{\mathcal{I},\text{pre}})$ given in Eq.~(\ref{Iconstraints}) in Sec.~\ref{SecSDP} such that $\text{tr}(\tilde{\chi}_{\mathcal{I},\text{pre}})=1$.

\section{Demonstrations of \textit{E}-preservability measures}
In general, \textit{E}-preservability measures can be used to analyze the entanglement preservation capability of all two-qubit processes that can be described by quantum operations \cite{Nielsen00,Chuang97}. In this section, we demonstrate the \textit{E}-preservability measures for the particular case of photonic and superconducting systems.

\subsection{Fusion of entangled photon pairs}\label{photon_fusion}
In photonic systems, the fusion of photon pairs \cite{Pan98,Pan12} superposes two individual photons in two different spatial modes ($1$ and $2$) at a polarizing beam splitter (PBS) and post-selects the two outputs in different modes ($a$ and $b$), as shown in Fig.~\ref{fusion}(a). The PBS transmits horizontal ($H$) polarization and reflects vertical ($V$) polarization. That is, after the PBS, the states $\ket{H_1H_2},$ $\ket{H_1V_2},$ $\ket{V_1H_2},$ $\ket{V_1V_2}$ become $\ket{H_bH_a},$ $\ket{H_bV_b},$ $\ket{V_aH_a},$ $\ket{V_aV_b}$, respectively, where ${H_i}$ (${V_i}$) denotes $H$ ($V$) polarization in the spatial mode $i=1,$ $2,$ $a,$ $b$. The post-selection operation in the different modes, $a$ and $b$, makes the fusion process of the photon pairs non-TP and the corresponding process matrix, denoted as $\tilde{\chi}_\text{fusion}$, is non-normalized.

$\tilde{\chi}_\text{fusion}$ can be represented in the form of quantum operations as follows \cite{Pan12}:
\begin{equation}\label{ideal_fusion}
\rho_\text{out}=\tilde{\chi}_\text{fusion}(\rho_\text{in})=M\rho_\text{in}M^\dagger,
\end{equation}
where
\begin{equation}\label{M}
M=\ket{H_{a}H_b}\!\!\bra{H_1H_2}+\ket{V_aV_b}\!\!\bra{V_1V_2},
\end{equation}
is the fusion operator.
Since $\rho_\text{in}=\ket{\phi^{+}}_{\text{in}}\!\!\tensor*[_{\text{in}}]{\bra{\phi^{+}}}{}$, the entangled state $\ket{\phi^{+}}_{\text{in}}=(\ket{H_1H_2}+\ket{V_1V_2})/\sqrt{2}$ remains an entangled state $\tilde{\chi}_\text{fusion}(\ket{\phi^{+}}_\text{in}\!\!\tensor*[_{\text{in}}]{\bra{\phi^{+}}}{})=\ket{\phi^{+}}_\text{out}\!\!\tensor*[_{\text{out}}]{\bra{\phi^{+}}}{}$ following the fusion operation, where $\ket{\phi^{+}}_\text{out}=(\ket{H_aH_b}+\ket{V_aV_b})/\sqrt{2}$. By contrast, $\rho_\text{in}$ is separable, and hence the state $\ket{+_1+_2}$, where $\ket{+_k}=(\ket{H_k}+\ket{V_k})/\sqrt{2}$ for $k=1,2$, becomes an entangled state $\ket{\phi^{+}}_\text{out}$ with the probability $1/2$, i.e., $\tilde{\chi}_\text{fusion}(\ket{+_1+_2}\!\!{\bra{+_1+_2}})=\ket{\phi^{+}}_\text{out}\!\!\tensor*[_{\text{out}}]{\bra{\phi^{+}}}{}/2$. Since the entangled state $\ket{\phi^{+}}_\text{in}\!\!\tensor*[_{\text{in}}]{\bra{\phi^{+}}}{}$ remains entangled after the fusion operation, the fusion process of the photon pairs $\tilde{\chi}_\text{fusion}$ is a capable process for entanglement preservation.

To quantify the \textit{E}-preservability of $\tilde{\chi}_\text{fusion}$ using $\alpha_\text{pre}$ and $\beta_\text{pre}$, we normalize the process matrix $\tilde{\chi}_\text{fusion}$ through $\chi_\text{fusion}=\tilde{\chi}_\text{fusion}/\text{tr}(\tilde{\chi}_\text{fusion})$. We note that $\alpha_\text{pre}$ and $\beta_\text{pre}$ of $\chi_\text{fusion}$ are both equal to $1$ and the \textit{E}-preservability criterion has a value of $F_{\mathcal{I},\text{pre}}=0.5$ for the photon fusion task.

The necessary condition for successful photon fusion is that the two photons must interfere at the PBS \cite{Pan12}. If the two photons do not arrive at the PBS simultaneously, their arrival times make them distinguishable after the PBS and the process becomes two projectors, namely $M_0=\ket{H_aH_b}\!\!\bra{H_1H_2}$ and $M_1=\ket{V_aV_b}\!\!\bra{V_1V_2}$. The resulting noise can be described as
\begin{equation}\label{chi_noise}
\rho_\text{out}=\tilde{\chi}_\text{noise}(\rho_\text{in})=M_0\rho_\text{in}M^\dagger_0+M_1\rho_\text{in}M^\dagger_1.
\end{equation}
Here, $\tilde{\chi}_\text{noise}$ is an incapable process since it cannot preserve entanglement and makes the entangled state $\ket{\phi^{+}}_\text{in}\!\!\tensor*[_{\text{in}}]{\bra{\phi^{+}}}{}$ become a separable state $(\ket{H_aH_b}\!\!\bra{H_aH_b}+\ket{V_aV_b}\!\!\bra{V_aV_b})/2$.

This imperfect photon fusion process can be described as
\begin{equation}\label{chi_Pnoise}
\tilde{\chi}_\text{expt}(p_\text{noise})=(1-p_\text{noise})\tilde{\chi}_\text{fusion}+p_\text{noise}\tilde{\chi}_\text{noise},
\end{equation}
where $p_\text{noise}$ is the noise intensity. The effect of noise on the $E$-preservability of the photon fusion process is illustrated in Fig.~\ref{fusion}(b). As shown, while $\chi_\text{expt}(p_\text{noise})$ are capable processes when $p_\text{noise}<1$, the \textit{E}-preservability decreases as the noise intensity $p_\text{noise}$ increases. The process matrices $\chi_\text{expt}$ for $p_\text{noise}=0$, $0.5$ and $1$ are shown in Fig.~\ref{fusion}(c) for illustration purposes. (See Appendix \ref{process_tomography} for details of the process matrix representation.)

\subsection{Quantum gates in IBM Q}\label{SecIBMQ}
This section examines the use of quantum gates to form a set of universal gates for quantum computation in IBM superconducting quantum computer (IBM Q) designated as \textit{ibmq\_5\_yorktown-ibmqx2} \cite{IBMQ}.
To quantify the \textit{E}-preservability of the single-qubit gates in IBM~Q, we first obtain the process matrices of these gates, denoted as $\chi^{(1)}_\text{expt}$, via QPT and then quantify the \textit{E}-preservability of $\chi_\text{expt}=\chi^{(1)}_\text{expt}\otimes\chi_I$ using the measures $\alpha_\text{pre}$ and $\beta_\text{pre}$ as shown in Fig.~\ref{main_concept}(b).
For ideal quantum gates, the \textit{E}-preservability measures have values of $\alpha_\text{pre}=1$ and $\beta_\text{pre}=1$, respectively, and the process fidelity threshold is $F_{\mathcal{I},\text{pre}}=0.5$. As shown in Table~\ref{tableone}, although all of the quantum gates can preserve entanglement, the \textit{E}-preservabilities of the single-qubit gates are greater than that of the two-qubit gate. An inspection of the measures $\alpha_\text{pre}$ and $\beta_\text{pre}$ in Table~\ref{tableone}, shows that IBM Q has high \textit{E}-preservability.

\begin{table}[t]
\caption{\textit{E}-preservability of quantum gates in IBM~Q. We implement seven quantum gates for universal quantum computing with IBM~Q and calculate the corresponding \textit{E}-preservability and the process fidelity $F_\text{expt}$. The process fidelities $F_\text{expt}$ of the identity gate ($I$), Pauli operators ($X$, $Y$, $Z$), Hadamard gate ($H$), $\pi/8$ gate ($T$) and CNOT gate are all greater than the process fidelity threshold $F_{\mathcal{I},\text{pre}}=0.5$. For ideal quantum gates, the \textit{E}-preservability measures are $\alpha_\text{pre}=1$ and $\beta_\text{pre}=1$.}
\begin{ruledtabular}
\begin{tabular}{ l c c c c c c c }
Measures & $I$ & $X$ & $Y$ & $Z$ & $H$ & $T$ & CNOT \\ \hline
 $\alpha_\text{pre}$ & $0.939$ & $0.970$ & $0.931$ & $0.930$ & $0.899$ & $0.875$ & $0.678$ \\
 $\beta_\text{pre}$ & $0.918$ & $0.960$ & $0.912$ & $0.904$ & $0.894$ & $0.868$ & $0.674$ \\
 $F_\text{expt}$ & $0.959$ & $0.980$ & $0.960$ & $0.953$ & $0.947$ & $0.934$ & $0.757$ \\  
\end{tabular}
\end{ruledtabular}
\label{tableone}
\end{table}

\section{Comparison between \textit{E}-preservability and entanglement creation capability}
To compare the \textit{E}-preservability of a process with the entanglement creation capability of the same process, we commence by reviewing the process capability of entanglement creation \cite{Kuo19}. A process is said to be an incapable process of entanglement creation, denoted as $\chi_{\mathcal{I},\text{cre}}$, if it cannot create entanglement from separable states and thus preserves the separability of the quantum system. Conversely, a process is said to be a capable process to create entanglement, if it cannot be described by $\chi_{\mathcal{I},\text{cre}}$. To make separable states remain separable states after an incapable process $\chi_{\mathcal{I},\text{cre}}$, the set of necessary constraints acting on the incapable process $D(\tilde{\chi}_{\mathcal{I},\text{cre}})$ are defined as
\begin{equation}
\tilde{\chi}_{\mathcal{I},\text{cre}}(\rho_{\text{in}})\geq0\ \forall \rho_{\text{in}};(\tilde{\chi}_{\mathcal{I},\text{cre}}(\rho_{\text{in}}))^{\text{PT}}\geq0\ \forall \rho_{\text{in}}\in s_{\text{sep}},\label{Dent}
\end{equation}
where $s_{\text{sep}}$ denotes the set of separable states.
The first constraint in Eq.~(\ref{Dent}) ensures that the output states are positive semi-definite for all the input states required in the QPT algorithm. The second constraint is based on the PPT criterion \cite{Peres96,Horodecki96}, and guarantees that, if the input states are separable states, the output states are separable states as well.

The entanglement creation capability \cite{Kuo19} of a process $\chi_{\text{expt}}$ can be measured through the capability composition, $\alpha_\text{cre}$, and capability robustness, $\beta_\text{cre}$. The former decomposes $\chi_{\text{expt}}$ into $\alpha_\text{cre}\chi_{\mathcal{C},\text{cre}}$ and $(1-\alpha_\text{cre})\chi_{\mathcal{I},\text{cre}}$, where $\chi_{\mathcal{C},\text{cre}}$ are capable processes with $\alpha_\text{cre}=1$; the latter considers the minimum noise that makes $\chi_{\text{expt}}$ become $\chi_{\mathcal{I},\text{cre}}$. In particular, $\alpha_\text{cre}$, can be used to quantify the portion of the process that cannot generate entanglement but can preserve entanglement of $\chi_{\text{expt}}$. First, according to the definition of capability composition of entanglement creation, $\chi_{\text{expt}}$ can be decomposed as
\begin{equation}
\chi_{\text{expt}}=\alpha_\text{cre}\chi_{\mathcal{C},\text{cre}}+(1-\alpha_\text{cre})\chi_{\mathcal{I},\text{cre}},\nonumber
\end{equation}
where $(1-\alpha_\text{cre})\chi_{\mathcal{I},\text{cre}}$ is the part of the process that cannot generate entangled states from separable states. To quantify the part of $\chi_{\text{expt}}$ that cannot generate entanglement but can preserve entanglement, we further use the \textit{E}-preservability composition to measure $(1-\alpha_\text{cre})\chi_{\mathcal{I},\text{cre}}$ and decompose it into two parts, i.e.,
\begin{equation}
(1-\alpha_\text{cre})\chi_{\mathcal{I},\text{cre}}=\alpha_\text{pre'}\chi_{\mathcal{C},\text{pre}}+(1-\alpha_\text{cre}-\alpha_\text{pre'})\chi_{\mathcal{I},\text{pre}}. \nonumber
\end{equation}
Thus, a process can be classified into three parts that are mutually exclusive, i.e.,
\begin{equation}
\chi_{\text{expt}}=\alpha_\text{cre}\chi_{\mathcal{C},\text{cre}}+\alpha_\text{pre'}\chi_{\mathcal{C},\text{pre}}+(1-\alpha_\text{cre}-\alpha_\text{pre'})\chi_{\mathcal{I},\text{pre}}, \nonumber
\end{equation}
where $\alpha_\text{cre}$ is the portion which can generate entanglement, $\alpha_\text{pre'}$ is the portion which cannot generate but can preserve entanglement, and $1-\alpha_\text{cre}-\alpha_\text{pre'}$ is the portion which can neither generate nor preserve entanglement.
Through Eqs.~(\ref{com}) and (\ref{COM}),
$\chi_{\text{expt}}$ can be represented as
\begin{equation}
\chi_{\text{expt}}=\alpha_\text{pre}\chi_{\mathcal{C},\text{pre}}+(1-\alpha_\text{pre})\chi_{\mathcal{I},\text{pre}}.\nonumber
\end{equation}
Comparing the above two equations and using the definitions of capable and incapable processes of \textit{E}-preservability,
we observe the following relation among these types of composition, $\alpha_{\text{pre}}=\alpha_{\text{cre}}+\alpha_{\text{pre'}}$. It is worth noting that the portion $\alpha_\text{pre'}\chi_{\mathcal{C},\text{pre}}+(1-\alpha_\text{cre}-\alpha_\text{pre'})\chi_{\mathcal{I},\text{pre}}$ can be non-normalized and less than 1.
Thus, one can obtain the portion $\alpha_\text{pre'}$ of $\chi_{\text{expt}}$ that cannot generate entanglement but can preserve entanglement from the \textit{E}-preservability composition and capability composition of entanglement generation as follows:
\begin{equation}
\alpha_\text{pre'}=\alpha_\text{pre}-\alpha_\text{cre}.\label{alpha_true_pre}
\end{equation}

In the following, we consider an example of two coupled qubits to demonstrate the \textit{E}-preservability and entanglement creation capabilities (See Fig.~\ref{main_concept}(c)). The Hamiltonian of coupled qubits with an intensity $h_{\text{int}}$ has the form \cite{Briegel01}
\begin{equation}
H_{\text{int}}=\frac{h_{\text{int}}}{2}\sum_{j,k=0}^1(-1)^{jk}\ket{jk}\!\bra{jk}.\label{haniltonian}
\end{equation}
The interaction between the qubits is equivalent to the quantum Ising model, which is an important primitive for creating cluster states in a one-way quantum computer \cite{Raussendorf01}. In considering the interaction with $H_{\text{int}}$, we assume that one of the qubits is depolarized at a rate $\gamma$. The master equation for the total system can then be expressed in the Lindblad form \cite{Breuer&Petruccione02,Hall14} as follows:
\begin{eqnarray}
\frac{\partial}{\partial t}\rho&=&-\frac{i}{\hbar}[H_{\text{int}},\rho]+\frac{\gamma}{2}[(I\otimes X)\rho (I\otimes X)\space \nonumber \\
&+&(I\otimes Y)\rho (I\otimes Y)+(I\otimes Z)\rho (I\otimes Z)\!\!-3\rho]
\end{eqnarray}
where $X$, $Y$, and $Z$ denote the Pauli-$X$, Pauli-$Y$ and Pauli-$Z$ matrices, respectively.

\begin{figure}[t]
\includegraphics[width=7cm]{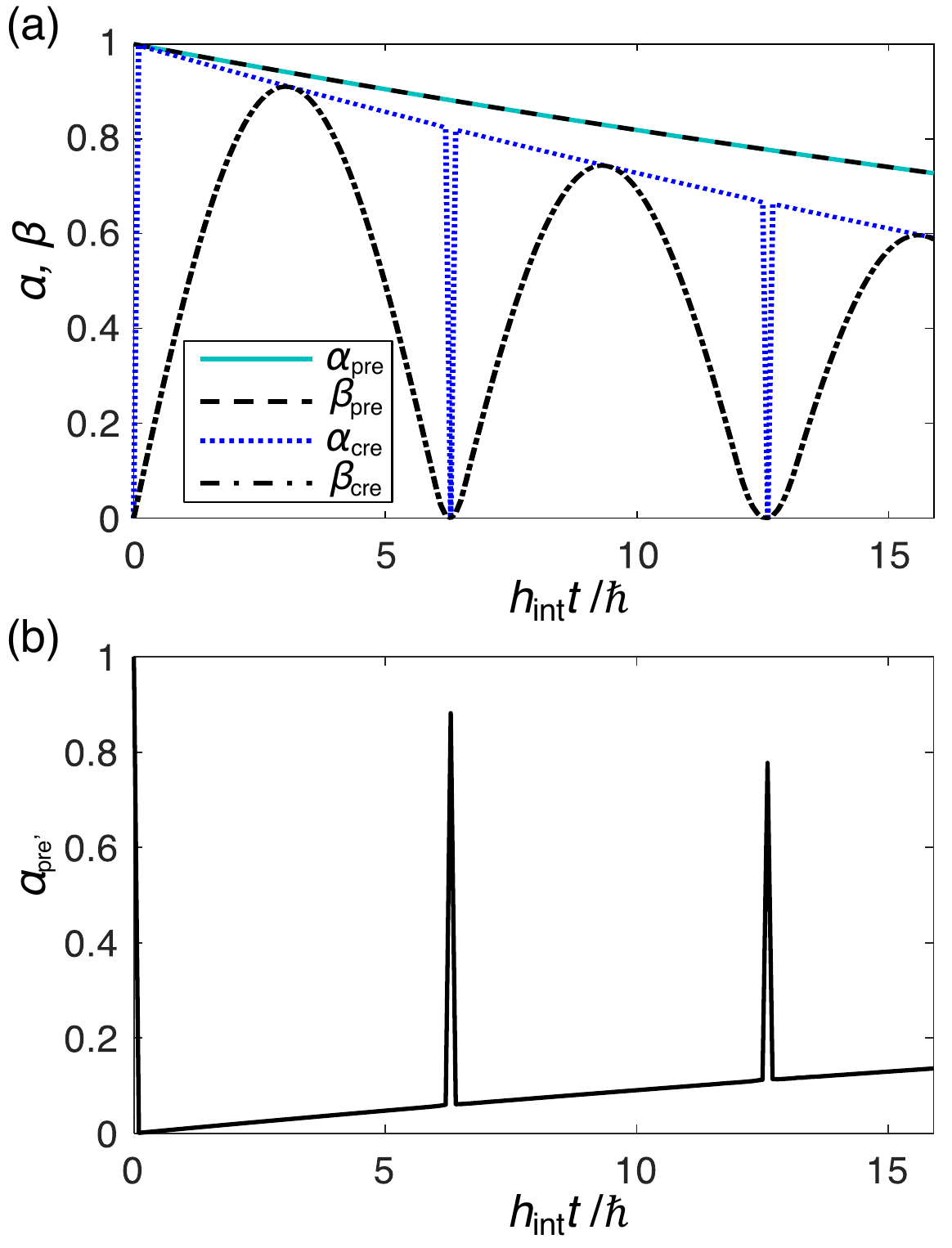}
\caption{Comparison between \textit{E}-preservability and process capability of entanglement creation. The capabilities in the dynamics $\chi_{\text{expt}}(t)$ of two coupled qubits under the Hamiltonian $H_{\text{int}}$ [Eq.~(\ref{haniltonian})] and a single-qubit depolarizing channel are examined (a) using the composition measures $\alpha_\text{pre}$, $\alpha_\text{cre}$ and the robustness measures $\beta_\text{pre}$, $\beta_\text{cre}$, and (b) through the composition $\alpha_\text{pre'}$ of the process that cannot generate but can preserve entanglement in Eq.~(\ref{alpha_true_pre}). The depolarizing rate $\gamma$ affects the curves of $\alpha$ and $\beta$, and $\gamma$ is equal to $0.02$, in the considered example.}
\label{dynamics}
\end{figure}

Figure~\ref{dynamics}(a) shows the variations of the composition and robustness of \textit{E}-preservability, $\alpha_\text{pre}$ and $\beta_\text{pre}$, and the process capability of entanglement creation, $\alpha_\text{cre}$ and $\beta_\text{cre}$. Note that the depolarization rate is set as $\gamma=0.02$. For $h_{\text{int}}t/\hbar=0$, the process is an identity transform process which cannot create entanglement from separable states but can preserve entanglement. Consequently, the capability measures of entanglement creation are equal to 0 and the \textit{E}-preservability measures have their maximum values $\alpha_\text{pre}=\beta_\text{pre}=1$. When $h_{\text{int}}t/\hbar=\pi$, the process is close to a controlled-$Z$ gate, for which the unitary transform is $U_\text{CZ}=\sum_{j,k=0}^1(-1)^{jk}\ket{jk}\!\bra{jk}$. In other words, the process can create and preserve entanglement, and hence the capability measures of entanglement creation have their maximum values. However, the \textit{E}-preservability measures have values of $\alpha_\text{pre}=\beta_\text{pre}=0.94$, which are lower than those of the process at $h_{\text{int}}t/\hbar=0$ due to the affects of the depolarizing channel. For $h_{\text{int}}t/\hbar=2n\pi,$ $n=0,1,2...$, the processes have the form of local operations on each qubit that are thus incapable of entanglement creation but capable to preserve entanglement and have non-zero $\alpha_\text{pre'}$ as shown in Fig.~\ref{dynamics} (b).

It is worth noting that the values of the \textit{E}-preservability are always greater than those of the process capability of entanglement creation. Furthermore, for a process with an entanglement creation capability, there should exist at least one output state that is entangled, and hence the process must also have \textit{E}-preservability.

For illustrative process of entangled photon pair fusion considered in Sec.~\ref{photon_fusion}, the measures of the process capability of entanglement creation, i.e., $\alpha_\text{cre}$ and $\beta_\text{cre}$, have the same variations as those of the \textit{E}-preservability shown in Fig.~\ref{fusion}(b). For the quantum gates in IBM~Q, all of the single-qubit gates are local operations, and hence do not have the capability of entanglement creation, with $\alpha_\text{pre}=\alpha_\text{pre'}$. By contract, the CNOT gate that couples the two qubits shows the process capability of entanglement creation, i.e., $\alpha_\text{cre}=0.6745$ and $\beta_\text{cre}=0.5996$, and hence has a less quantity $\alpha_\text{pre'}=0.0035$. It is worth noting that, in these examples, $\alpha_\text{cre}$ and $\beta_\text{cre}$ are smaller than $\alpha_\text{pre}$ and $\beta_\text{pre}$, respectively. In general, the examples show that if a process can create entanglement, it also has the capability to preserve entanglement. Therefore, non-zero \textit{E}-preservability is a precondition for a process used to create entanglement.

\section{Conclusion and outlook}
In conclusion, we have considered a method for characterizing the entanglement preservation capability of a process. Notably, the method can be used for both CPTP and non-TP CP processes and does not require any ancillary system. The used method has shown that entanglement preservability can be considered as a quantum process capability and completes quantum process capability theory by quantifying both entanglement creation and entanglement preservation. We have shown that one can use experimentally feasible methods without using entangled states, i.e., through the quantum process tomography algorithm, to quantify the entanglement preservation capability of all processes that can be described using the general theory of quantum operations. We have additionally discussed the relationship between entanglement preservability and the capability of entanglement creation in quantum process capability theory and have shown that entanglement preservability is a necessary precondition for creating entanglement. Moreover, since \textit{E}-preservability composition and robustness satisfy the properties of a proper measure, they can faithfully quantify \textit{E}-preservability in practical experiments, such as creating entangled photon pairs through photon fusion.

To completely characterize physical processes via quantum process tomography, it is necessary to acquire many more measurement settings than just performing state tomography on specific states. Future studies will aim to address this problem for the illustrative case of entanglement preservability by devising methods other than quantum process tomography for characterization purposes. Since photon fusion can generate multipartite entanglement, entanglement preservability might be used to examine the generation of multipartite entanglement by evaluating the processes of photon fusion between different entangled pairs. Besides entanglement, there are many other quantum correlations that may also be usefully considered for quantum-information tasks in different situations. Furthermore, our method may potentially be extended to the characterization of other quantum correlations, e.g., Einstein-Podolsky-Rosen \cite{Wiseman07} steering or Bell nonlocality \cite{Brunner14}, in dynamical processes.

\section*{ACKNOWLEDGMENTS}
This work is partially supported by the Ministry of Science and Technology, Taiwan, under Grant Number MOST 107-2628-M-006-001-MY4.

\appendix

\section{Quantum process tomography and process matrix}\label{process_tomography}

\subsection{Basic concept}
According to the quantum operations formalism \cite{Nielsen00,Chuang97}, the input states $\rho_\text{in}$ and output states $\rho_\text{out}$ of the unknown dynamics of a quantum system can be associated via the following dynamical mapping:
\begin{eqnarray}
\chi:\rho_\text{in}\mapsto \rho_\text{out}.\nonumber
\end{eqnarray}
That is, the output state of an $n$-qubit system can be explicitly represented as
\begin{equation}
\rho_\text{out}\equiv\chi(\rho_\text{in})=\sum_{k=1}^{4^n}\sum_{j=1}^{4^n}\chi_{kj}E_{k}\rho_\text{in}E^{\dag}_{j},\label{rhoout}
\end{equation}
where
\begin{equation}
E_{k}=\bigotimes_{m=1}^{n}\ket{k_{m}}\!\!\bra{k_{m+n}},\label{ek}
\end{equation}
and $k=1+\sum_{i=1}^{2n}k_i2^{i-1}$ for $k_i\in\{0,1\}$. To determine the coefficients $\chi_{kj}$ which constitute the so-called process matrix, $\chi$, we consider the following $4^{n}$ inputs:
\begin{equation}
\rho_{\text{in},k'}=E_{k'}=\bigotimes_{m=1}^{n}\ket{k'_{m}}\!\!\bra{k'_{m+n}},\nonumber
\end{equation}
for $k'=1,2,...,4^{n}$. From Eq.~(\ref{rhoout}), we obtain the corresponding outputs as
\begin{equation}
\chi(\rho_{\text{in},k'})\!=\!\sum_{k_{1}=0}^{1}\!...\!\sum_{j_{n}=0}^{1}\bigotimes_{m=1}^{n}\ket{k_{m}}\!\!\bra{j_{m}}\chi_{p(\bold{k},k')q(\bold{j},k')},\label{chiout}
\end{equation}
where $k'=1+\sum_{i=1}^{2n}k'_i2^{i-1}$ for $k'_i\in\{0,1\}$, $\bold{k}=(k_{1},...,k_{n})$, $\bold{j}=(j_{1},...,j_{n})$,
\begin{equation}
p(\bold{k},k')=1+\sum_{i=1}^{n}k_{i}2^{i-1}+\sum_{i=1}^{n}k'_{i}2^{n+i-1},
\end{equation}
and
\begin{equation}
q(\bold{j},k')=1+\sum_{i=1}^{n}j_{i}2^{i-1}+\sum_{i=n+1}^{2n}k'_{i}2^{i-1}.
\end{equation}
As the output $\chi(\rho_{\text{in},k'})$ is determined using quantum state tomography, we have full knowledge of the output matrix:
\begin{eqnarray}
\rho_{\text{out},k'}&=&\chi(\rho_{\text{in},k'})\nonumber\\
&=&\sum_{k_{1}=0}^{1}...\sum_{j_{n}=0}^{1}\bigotimes_{m=1}^{n}\ket{k_{m}}\!\!\bra{j_{m}}\rho^{(k')}_{\bold{k}\bold{j}},\label{qstout}
\end{eqnarray}
i.e., all the $2^n\times2^n=4^n$ matrix elements $\rho^{(k')}_{\bold{k}\bold{j}}$ are determined. By comparing Eq.~(\ref{chiout}) with Eq.~(\ref{qstout}), one can obtain the process matrix $\chi$ with the $4^n\times4^n$ matrix elements as
\begin{equation}
\chi_{p(\bold{k},k')q(\bold{j},k')}=\rho^{(k')}_{\bold{k}\bold{j}}.
\end{equation}

\subsection{Illustrative examples}

\subsubsection{Single-qubit process tomography}
For the dynamical process of a single qubit, $n=1$, we have the following four operators $E_{k}$ according to Eq.~(\ref{ek}): $E_{1}=\ket{0}\!\!\bra{0}$, $E_{2}=\ket{1}\!\!\bra{0}$, $E_{3}=\ket{0}\!\!\bra{1}$, and $E_{4}=\ket{1}\!\!\bra{1}$. As shown in Eqs.~(\ref{chiout}) and (\ref{qstout}), the output matrices of $E_{k'}$ have the forms
\begin{eqnarray}
\rho_{\text{out},1}\!&=&\chi_{11}\ket{0}\!\!\bra{0}+\chi_{12}\ket{0}\!\!\bra{1}+\chi_{21}\ket{1}\!\!\bra{0}+\chi_{22}\ket{1}\!\!\bra{1}\nonumber\\
&=&\left[\begin{array}{cc}\chi_{11} & \chi_{12} \\ \chi_{21} & \chi_{22}\end{array}\right],\nonumber\\
\rho_{\text{out},3}&=&\chi_{13}\ket{0}\!\!\bra{0}+\chi_{14}\ket{0}\!\!\bra{1}+\chi_{23}\ket{1}\!\!\bra{0}+\chi_{24}\ket{1}\!\!\bra{1}\nonumber\\
&=&\left[\begin{array}{cc}\chi_{13} & \chi_{14} \\ \chi_{23} & \chi_{24}\end{array}\right],\nonumber\\
\rho_{\text{out},2}&=&\chi_{31}\ket{0}\!\!\bra{0}+\chi_{32}\ket{0}\!\!\bra{1}+\chi_{41}\ket{1}\!\!\bra{0}+\chi_{42}\ket{1}\!\!\bra{1}\nonumber\\
&=&\left[\begin{array}{cc}\chi_{31} & \chi_{32} \\ \chi_{41} & \chi_{42}\end{array}\right],\nonumber\\
\rho_{\text{out},4}&=&\chi_{33}\ket{0}\!\!\bra{0}+\chi_{34}\ket{0}\!\!\bra{1}+\chi_{43}\ket{1}\!\!\bra{0}+\chi_{44}\ket{1}\!\!\bra{1}\nonumber\\
&=&\left[\begin{array}{cc}\chi_{33} & \chi_{34} \\ \chi_{43} & \chi_{44}\end{array}\right].\nonumber
\end{eqnarray}
It is clear that once the density matrices of these four outputs are known by using quantum state tomography, the coefficients $\chi_{kj}$ can be determined. It is worth noting that since the coherence terms can be decomposed as $E_2=\ket{+}\!\!\bra{+}\!-\!i\ket{R}\!\!\bra{R}\!-\!\hat{I}^{\dag}$ and $E_3=\ket{+}\!\!\bra{+}\!+\!i\ket{R}\!\!\bra{R}\!-\!\hat{I}$, where $\ket{+}=(\ket{0}+\ket{1})/\sqrt{2}$, $\ket{R}=(\ket{0}+i\ket{1})/\sqrt{2}$, and $\hat{I}=e^{i\pi/4}I/\sqrt{2}$, the output states are experimentally obtainable by measuring the density matrices of $\rho_{\text{out},+}=\chi(\ket{+}\!\!\bra{+})$ and $\rho_{\text{out},R}=\chi(\ket{R}\!\!\bra{R})$. Therefore, with the above results, we arrive at
\begin{eqnarray}
\chi
&=&\frac{1}{2}\left[\begin{array}{cccc}\chi_{11} & \chi_{12} & \chi_{13} & \chi_{14} \\\chi_{21} & \chi_{22} & \chi_{23} & \chi_{24} \\\chi_{31} & \chi_{32} & \chi_{33} & \chi_{34} \\\chi_{41} & \chi_{42} & \chi_{43} & \chi_{44}\end{array}\right]\nonumber\\
&=&\frac{1}{2}\left[\begin{array}{cc}\rho_{\text{out},1} & \rho_{\text{out},+}+i\rho_{\text{out},R}-\tilde{I}_\text{out} \\ \rho_{\text{out},+}-i\rho_{\text{out},R}-\tilde{I}_\text{out}^{\dag} & \rho_{\text{out},4}\end{array}\right],\nonumber\\
\end{eqnarray}
where $\tilde{I}_\text{out}=e^{i\pi/4}(\rho_{\text{out},1}+\rho_{\text{out},4})/\sqrt{2}$, and the factor of $1/2$ is a normalization constant set such that $\chi_{\text{expt}}$ can be treated as a density matrix.

\subsubsection{Two-qubit process tomography}
For the dynamical process of two qubits, $n=2$, e.g., the process matrices shown in Fig.~\ref{fusion}(c), $E_{k}$ have the form
\begin{equation}\label{Ek_two}
E_{k}=\ket{k_{1}k_{2}}\!\!\bra{k_{3}k_{4}}
\end{equation}
according to Eq.~(\ref{ek}), and $E_{1}=\ket{00}\!\!\bra{00}$, $E_{2}=\ket{10}\!\!\bra{00}$, $E_{3}=\ket{01}\!\!\bra{00}$, ..., and $E_{16}=\ket{11}\!\!\bra{11}$.
To use experimentally preparable input states to obtain process matrices, we decompose $E_{k}$, $k=1,2,...,16$, as a linear combination of the following density matrices of the states $\rho'_{\text{in},m}=\ket{m}\!\!\bra{m}, \ket{m}\in$ $\{\ket{00}$, $\ket{01}$, $\ket{10}$, $\ket{11}$, $\ket{0+}$, $\ket{0R}$, $\ket{1+}$, $\ket{1R}$, $\ket{+1}$, $\ket{+0}$, $\ket{R1}$, $\ket{R0}$, $\ket{\phi^+}$, $\ket{\phi^{+i}}$, $\ket{\psi^+}$, $\ket{\psi^{+i}}\}$, where $\ket{\phi^+}=(\ket{00}+\ket{11})/\sqrt{2}$, $\ket{\phi^{+i}}=(\ket{00}+i\ket{11})/\sqrt{2}$, $\ket{\psi^+}=(\ket{01}+\ket{10})/\sqrt{2}$, and $\ket{\psi^{+i}}=(\ket{01}+i\ket{10})/\sqrt{2}$. The output matrix of $E_{k'}$, denoted by $\rho_{\text{out},k'}$, thus can be represented by using the output density matrices of these states, denoted as $\rho'_{\text{out},m}=\chi(\rho'_{\text{in},m})$. That is,
\begin{eqnarray}
\rho_{\text{out},1}&=&\rho'_{\text{out},00\nonumber},\\
\rho_{\text{out},2}&=&\rho'_{\text{out},0+}+i\rho'_{\text{out},0R}-\frac{e^{i\pi/4}}{\sqrt{2}}(\rho'_{\text{out},00}+\rho'_{\text{out},01}),\nonumber\\
\rho_{\text{out},3}&=&\rho'_{\text{out},+0}+i\rho'_{\text{out},R0}-\frac{e^{i\pi/4}}{\sqrt{2}}(\rho'_{\text{out},00}+\rho'_{\text{out},10}),\nonumber\\
\rho_{\text{out},4}&=&\rho'_{\text{out},\phi^+}+i\rho'_{\text{out},\phi^{+i}}-\frac{e^{i\pi/4}}{\sqrt{2}}(\rho'_{\text{out},00}+\rho'_{\text{out},11}),\nonumber\\
\rho_{\text{out},5}&=&\rho'_{\text{out},0+}-i\rho'_{\text{out},0R}-\frac{e^{-i\pi/4}}{\sqrt{2}}(\rho'_{\text{out},00}+\rho'_{\text{out},01}),\nonumber\\
\rho_{\text{out},6}&=&\rho'_{\text{out},01},\nonumber\\
\rho_{\text{out},7}&=&\rho'_{\text{out},\psi^+}+i\rho'_{\text{out},\psi^{+i}}-\frac{e^{i\pi/4}}{\sqrt{2}}(\rho'_{\text{out},01}+\rho'_{\text{out},10}),\nonumber\\
\rho_{\text{out},8}&=&\rho'_{\text{out},+1}+i\rho'_{\text{out},R1}-\frac{e^{i\pi/4}}{\sqrt{2}}(\rho'_{\text{out},01}+\rho'_{\text{out},11}),\nonumber\\
\rho_{\text{out},9}&=&\rho'_{\text{out},+0}+i\rho'_{\text{out},R0}-\frac{e^{i\pi/4}}{\sqrt{2}}(\rho'_{\text{out},00}+\rho'_{\text{out},10}),\nonumber\\
\rho_{\text{out},10}&=&\rho'_{\text{out},\psi^+}-i\rho'_{\text{out},\psi^{+i}}-\frac{e^{-i\pi/4}}{\sqrt{2}}(\rho'_{\text{out},01}+\rho'_{\text{out},10}),\nonumber\\
\rho_{\text{out},11}&=&\rho'_{\text{out},10},\nonumber\\
\rho_{\text{out},12}&=&\rho'_{\text{out},1+}+i\rho'_{\text{out},1R}-\frac{e^{i\pi/4}}{\sqrt{2}}(\rho'_{\text{out},10}+\rho'_{\text{out},11}),\nonumber\\
\rho_{\text{out},13}&=&\rho'_{\text{out},\phi^+}-i\rho'_{\text{out},\phi^{+i}}-\frac{e^{-i\pi/4}}{\sqrt{2}}(\rho'_{\text{out},00}+\rho'_{\text{out},11}),\nonumber\\
\rho_{\text{out},14}&=&\rho'_{\text{out},+1}-i\rho'_{\text{out},R1}-\frac{e^{-i\pi/4}}{\sqrt{2}}(\rho'_{\text{out},01}+\rho'_{\text{out},11}),\nonumber\\
\rho_{\text{out},15}&=&\rho'_{\text{out},1+}-i\rho'_{\text{out},1R}-\frac{e^{-i\pi/4}}{\sqrt{2}}(\rho'_{\text{out},10}+\rho'_{\text{out},11}),\nonumber\\
\rho_{\text{out},16}&=&\rho'_{\text{out},11}.\label{PT_twoQ}
\end{eqnarray}
These 16 output states $\rho'_{\text{out},m}$ are sufficient to describe a process that makes all input states separable as shown in Sec.~\ref{properties}.

Taking $\rho_{\text{in},k'}=E_{9}=\ket{10}\!\!\bra{00}$ for example purposes, the output matrix of $E_{9}$ in Eq.~(\ref{qstout}) contains some of the process matrix elements in Eq.~(\ref{chiout}) and has the form
\begin{eqnarray}
\rho_{\text{out},9}\!=\left[\begin{array}{cccc}\chi_{1,9} & \chi_{1,10} & \chi_{1,11} & \chi_{1,12} \\\chi_{2,9} & \chi_{2,10} & \chi_{2,11} & \chi_{2,12} \\\chi_{3,9} & \chi_{3,10} & \chi_{3,11} & \chi_{3,12} \\\chi_{4,9} & \chi_{4,10} & \chi_{4,11} & \chi_{4,12}\end{array}\right].
\end{eqnarray}
Since $E_{9}$ can be decomposed as $E_9=\ket{+0}\!\!\bra{+0}\!+\!i\ket{R0}\!\!\bra{R0}\!-\!e^{i\pi/4}(\ket{00}\!\!\bra{00}\!+\!\ket{10}\!\!\bra{10})/\sqrt{2}$, one can obtain $\rho_{\text{out},9}$ by measuring the density matrices of $\rho'_{\text{out},00}=\chi(\ket{00}\!\!\bra{00})$, $\rho'_{\text{out},10}=\chi(\ket{10}\!\!\bra{10})$, $\rho'_{\text{out},+0}=\chi(\ket{+0}\!\!\bra{+0})$, and $\rho'_{\text{out},R0}=\chi(\ket{R0}\!\!\bra{R0})$ as shown in Eq.~(\ref{PT_twoQ}).
With $\rho_{\text{out},k'}$ for $k'=1,2,...,16$, the process matrix, $\chi$, can be obtained as
\begin{eqnarray}
\chi\!=\frac{1}{4}\left[\begin{array}{cccc}\rho_{\text{out},1} & \rho_{\text{out},5} & \rho_{\text{out},9} & \rho_{\text{out},13} \\\rho_{\text{out},2} & \rho_{\text{out},6} & \rho_{\text{out},10} & \rho_{\text{out},14} \\\rho_{\text{out},3} & \rho_{\text{out},7} & \rho_{\text{out},11} & \rho_{\text{out},15} \\\rho_{\text{out},4} & \rho_{\text{out},8} & \rho_{\text{out},12} & \rho_{\text{out},16}\end{array}\right],
\end{eqnarray}
where the constant $1/4$ is a normalization factor.

In the fusion process, $\tilde{\chi}_\text{fusion}$, in Eq.~(\ref{ideal_fusion}), the fusion operator, $M$, can be represented by $E_{k}$ [Eq.~(\ref{Ek_two})] as $M=\ket{H_aH_b}\!\!\bra{H_1H_2}+\ket{V_aV_b}\!\!\bra{V_1V_2}=\ket{00}\!\!\bra{00}+\ket{11}\!\!\bra{11}=E_{1}+E_{16}$, where $\ket{H_i}\equiv\ket{0}$ and $\ket{V_i}\equiv\ket{1}$ for $i=0,1,a,b$. According to Eq.~(\ref{rhoout}), the entries of $\tilde{\chi}_\text{fusion}$ are $\chi_{1,1}=\chi_{1,16}=\chi_{16,1}=\chi_{16,16}=1/4$, with the other elements all equal to zero.
The normalized process matrix of $\tilde{\chi}_\text{fusion}$, i.e. ${\chi}_\text{fusion}$, is shown in Fig.~\ref{fusion}(c) with $p_\text{noise}=0$. Similarly, the $\tilde{\chi}_\text{noise}$ in Eq.~(\ref{chi_noise}) can be represented by a process matrix with elements of $\chi_{1,1}=\chi_{16,16}=1/4$ and $\chi_{kj}=0$ otherwise, since $M_0$ and $M_1$ are $E_{1}$ and $E_{16}$ in Eq.~(\ref{Ek_two}), respectively. The normalized process matrix, ${\chi}_\text{noise}$, is shown in Fig.~\ref{fusion}(c) with $p_\text{noise}=1$. For the imperfect photon fusion process [Eq.~(\ref{chi_Pnoise})], the process matrix, $\tilde{\chi}_\text{expt}(p_\text{noise})$, is a linear combination of $\tilde{\chi}_\text{fusion}$ and $\tilde{\chi}_\text{noise}$. For $p_\text{noise}=0.5$, the entries of $\tilde{\chi}_\text{expt}(0.5)$ are $\chi_{1,1}=\chi_{16,16}=1/4$, $\chi_{1,16}=\chi_{16,1}=1/8$ with all the other elements equal to zero. The corresponding process matrix, ${\chi}_\text{expt}(0.5)$, is shown in Fig.~\ref{fusion}(c) with $p_\text{noise}=0.5$.\\

\section{Comparison between \textit{E}-preservability and quantifiers in channel resource theory}\label{ComparisonRT}
Both channel resource theory and QPC theory fit the goal of resource theories \cite{Gonda19} to divide objects, e.g., states, channels, or processes, into free and non-free sets. Here, to discuss the relationships of these two theories, we point out their differences from three different aspects.

In channel resource theories, two methods are available to quantify the ability of a process to preserve entanglement: recourse preservability theory \cite{Hsieh19} and quantum memory quantification theory in Ref.~\cite{Yuan19}. As considering entanglement as a resource, resource preservability theory reflects the entanglement preservability. This section of the paper discusses the relation between the \textit{E}-preservability theory considered in this work and existing resource preservability theory ~\cite{Hsieh19} reported in the literature from three perspectives, namely purpose, definition and application.

(i) Purpose. Resource preservability theory \cite{Hsieh19} aims to be constructed as a complete mathematical structure extended from state resource theory. In particular, it inherits three ingredients from state resource theory: resource channels, free channels, and free super-channels. According to resource theory, resource preservability theory focuses on free operations of a state resource and classifies these operations into resource channels (i.e., the free operations that preserve the state resource) and free channels (i.e., the free operations that completely destroy the state resource for every input).

Free super-channels are free operations in channel resource theories that map one free channel into another. In interpreting super-channels, interactions between the main system and ancillary systems are included. Therefore, in practical experiments, if one cannot get and use the optimized ancillary systems needed to measure the resource preservability, theoretical ancillary systems are used for obtaining the upper bound of the realistic resource preservability quantities. Through resource preservability theory, one can analyze resource preservability with a complete mathematical structure extended from state resource theory. By contrast, QPC theory offers a method with which one can analyze \textit{E}-preservability directly through the primitive measurement results of the output states.

(ii) Definition. A process shows non-zero resource preservability \cite{Hsieh19} to entanglement if the process maintains or partially degrades the entanglement quantity of at least one input entangled state, e.g. the entanglement quantified in max-relative entropy. For a process that is not a free operation (i.e., the process can generate entanglement from separable states) or not a TP process (i.e., not a channel \cite{Hsieh19}), resource preservability theory cannot be used to analyze the resource preservability of the process. Compared to resource preservability theory, however, QPC theory relaxes the constraints imposed on the expectations of a state resource, i.e., a state resource cannot be created under free operations. Therefore, using QPC theory, a process that has the ability to create entanglement can be characterized as a process having \textit{E}-preservability.

(iii) Application. Since resource preservability theory is not applicable to processes that can generate state resources \cite{Hsieh19}, if one wishes to objectively characterize resource preservability in an experimental process, it is necessary to check whether the process is a free operation. A process that is not a free operation cannot be characterized using resource preservability theory. However, in QPC theory, the \textit{E}-preservability of a process can be obtained directly using the primitive measurement results.

While there exist undoubted differences between resource preservability theory and QPC theory, as described above, a comprehensive comparison and investigation into the precise relationship between them is still lacking. Nonetheless, both theories can be used to quantify entanglement experiments with their own quantifiers satisfying the properties needed for a proper measure.

Another quantifier based on channel resource theory is the quantum memory quantifier proposed in \cite{Yuan19}. The authors in this study considered the ability of quantum memories to store quantum information as a physical resource. Therefore, according to channel resource theory, they defined free channels as entanglement breaking channels. Furthermore, a free transformation that maps entanglement breaking channels into entanglement breaking channels was defined as a free super-channel.

The quantifier considered in Ref.~\cite{Yuan19} can be used to quantify entanglement preservability since entanglement breaking channels are considered to be free channels and directly reflect the ability of a single-qubit channel to preserve entanglement. However, the quantifier can only be used to analyze the preservability of single-qubit channels since entanglement breaking channels are single-qubit channels. By contrast, in the \textit{E}-preservability measure considered in the present work, incapable processes are regarded as entanglement-annihilating channels, as defined in Ref.~\cite{Moravcikova10}, which act on two qubits and make all output states separable. Thus, the \textit{E}-preservability measure can be used to quantify all two-qubit processes.


\begin{references}
\bibitem{example_LOCC} Here, we propose a concrete example that demonstrates Eq.~(\ref{ChiPreI}) through LOCC. In the considered scenario, Alice first measures her local qubit with projectors $|{0_A}\rangle\langle{0_A}|$ and $|{1_A}\rangle\langle{1}_A|$. She then sends the outcome via classical communication to Bob, who prepares a local state conditioned on Alice's measurement result. This can be realized via local operations, i.e., measurement and preparation, as $\chi_\text{expt}(\rho_\text{in}) = |{0_A0_B}\rangle\langle{0_A0_B}|\text{tr}(|{0_A}\rangle\langle{0_A}|\otimes\mathbb{I}_B
\rho) + |{1_A1_B}\rangle\langle{1_A1_B}|\text{tr}(|{1_A}\rangle\langle{1_A}|\otimes\mathbb{I}_B \rho)$. Thus, the example can be written as $\chi_{\mathcal{I},{\rm pre}}=1/2\chi^A_1\otimes\chi^B_1+1/2\chi^A_2\otimes\chi^B_2$, where $\chi^A_1$ and $\chi^A_2$ are process matrices of projectors $|{0_A}\rangle\langle{0_A}|$ and $|{1_A}\rangle\langle{1_A}|$, respectively, and $\chi^B_1$ and $\chi^B_2$ are preparation processes that respectively output $|{0_B}\rangle\langle{0_B}|$ and $|{1_B}\rangle\langle{1_B}|$.
\end{references}
\end{document}